\documentclass[aip,pre,reprint,superscriptaddress]{revtex4-2}
\usepackage{amsmath}
\usepackage{graphicx}
\usepackage{bm}
\usepackage{dcolumn}

\usepackage{comment} 
\usepackage{color}
\usepackage{ulem}
\usepackage{enumerate}
\usepackage[colorlinks,
linkcolor=blue, urlcolor=blue, anchorcolor=blue, citecolor=blue]{hyperref}
\begin{document}
	
	
	\title{Evolution of the rippled inner-interface-initiated ablative Rayleigh-Taylor instability in laser-ablating high-Z doped targets}
	
	\author{W. Xiong}
	\affiliation{Department of Nuclear Science and Technology, National University of Defense Technology, Changsha 410073, China}
	
	\author{X. H. Yang}\thanks{Electronic mail: xhyang@nudt.edu.cn}
	\affiliation{Department of Nuclear Science and Technology, National University of Defense Technology, Changsha 410073, China}
	
	\author{Z. H. Chen}
	\affiliation{Department of Nuclear Science and Technology, National University of Defense Technology, Changsha 410073, China}
	
	\author{B. H. Xu}
	\affiliation{Department of Nuclear Science and Technology, National University of Defense Technology, Changsha 410073, China}
	
	\author{Z. Li}
	\affiliation{Department of Nuclear Science and Technology, National University of Defense Technology, Changsha 410073, China}
	
	\author{B. Zeng}
	\affiliation{Department of Nuclear Science and Technology, National University of Defense Technology, Changsha 410073, China}
	
	\author{G. B. Zhang}\thanks{Electronic mail: zgb830@163.com}
	\affiliation{Department of Nuclear Science and Technology, National University of Defense Technology, Changsha 410073, China}
	
	\author{Y. Y. Ma}
	\affiliation{School of Automation and Electronic Information, Xiangtan University, Xiangtan 411105, China}
	\affiliation{School of Physics and Electronics, Hunan University, Changsha 410082, China}

	
	\date{\today}
	
	\begin{abstract}
		Rippled interface between the ablator and DT ice can feedout and form the perturbation seeds for the ablative Rayleigh-Taylor (ART) instability, which negatively affects direct-drive inertial confinement fusion (ICF). 
		However, the evolution of instability remains insufficiently studied, and the effect of high-Z dopant on it remains unclear.
		In this paper, we develop a theoretical model to calculate the feedout seeds and describe this instability. 
		Our theory suggests that the feedout seeds are determined by the ablation pressure and the adiabatic index, while the subsequent growth mainly depends on the ablation velocity.
		Two-dimensional radiation hydrodynamic simulations confirm our theory. It is shown that high-Z doped targets exhibit more severe feedout seeds, because of their higher ionization compared to undoped targets.
		However, the X-ray pre-ablation in high-Z doped targets significantly suppresses the subsequent growth, leading to the suppression of short-wavelength perturbations.
		But for long-wavelength perturbations, this suppression weakens, resulting in an increased instability in the high-Z doped targets.
		The results are helpful for understanding the inner-interface-initiated instability and the influence of high-Z dopant on it, providing valuable insights for target design and instability control in ICF.
	\end{abstract}
	
	\maketitle
	\section{Introduction}

	Recently, OMEGA has achieved a hot-spot fuel gain exceeding unity using the direct-drive ICF scheme, bringing direct-drive ignition closer to realization.\cite{OmegaRecent,OmegaExperiment01,OmegaExperiment02} To further increase energy gain, hydrodynamic instability must be controlled to an acceptable level. \cite{DTiceRoughnessControl1,DTiceRoughnessControl2} The instability typically arises from target defects, such as those on the inner or outer surface of ablator. Due to the production techniques of solid deuterium-tritium (DT) ice, defects inside the targets are generally rougher than those on the surface. \cite{InnerRougher,DTiceRoughness2,DTiceRoughness} These inner defects will couple to the ablation front through feedout dynamics and grow due to the ablative Richtmyer-Meshkov (ARM) and ART instabilities, ultimately leading to ablator-fuel mixing and implosion performance degradation. \cite{InnerReview,innersurfaceLei,WangLF01} Through the Defect-Induced Mix Experimental (DIME) campaigns \cite{DIME-1,DIME-2} and relevant experiments \cite{DIMEexperiments} on NIF and OMEGA, defects have been identified as a critical factor in ablator-fuel mixing and implosion yield reduction.\cite{MixtureByDefects,InnerExperiments,InnerExperiments2} As a result, many studies focused on instabilities initiated by inner defects.

	Due to their typical height of less than 10 $\mu$m, inner defects are difficult to observe in experiments. \cite{DTiceRoughness2,DTiceRoughnessControl3,DomeDefects}
	Therefore, research often relies on simulations and theoretical models.\cite{UseSimulation}
	One approach is to decompose these dome-shaped defects into different mode sinusoidal perturbations, namely the rippled perturbations, and analyze the evolution of each mode separately.\cite{ICFReview}

	Early research about the rippled inner-interface-initiated instability was mainly based on single-layer targets with a rippled rear surface. 
	Betti $et$ $al.$ \cite{FeedoutBetti} first establish a theoretical model describing this rippled rear-surface-initiated instability at long-wavelength ($kd_{ps}<1$, where $d_{ps}$ and $k$ denote the postshock target thickness and the perturbation wave number). The asymptotic solution of the perturbation amplitude can be approximated by
	$\Delta_{\mathrm{fs}}\approx1/4[\Delta_0/(kd_0)+\sqrt{3\rho_{\mathrm{ps}}/(5\rho_0)}\Delta_r/\sqrt{kd_0}]\times\exp[\int_{t_\mathrm{rb}}^t\sqrt{kg_\mathrm{fs}(t^{\prime})}dt^{\prime}]$,
	where $\Delta_0$ and $\Delta_r$ denote the amplitude of the initial rippled inner interface and the reflected rarefaction wave. 
	The evolution of short-wavelength ($kd_{ps}>1$) perturbations is slightly different from that of long-wavelength perturbations. 
	Reflected rarefaction wave results in lower pressure at the perturbation valley, driving a lateral mass flow from the perturbation peak to the valley.\cite{FeedoutAlexander} 
	Therefore, the areal-mass perturbation exhibits the damped oscillation, which can be described by $\delta m/\delta m_0 =4\sqrt{2}[\gamma-1+\sqrt{2\gamma(\gamma-1)}]/(\gamma-1)^2\times \sin \Omega\tau/\Omega\tau$, where $\gamma$ is the adiabatic exponent, or the specific heat ratio. \cite{RearSurfaceAlexenzder, RarefactionAlexenzder,FeedoutAglitskiy}
	The validity of above theories has been confirmed by subsequent experiments. \cite{FeedoutShigmori,ARTjGrun,FeedoutExperimentsAglitskiy,FeedoutExperimentsAglitskiy02}
	
	Considering the density difference between the ablator (CH) and DT ice is limited (generally about five times), the perturbation evolution in single-layer targets significantly differs from that in multi-layer targets, resulting previous theories fail to accurately describe. \cite{FeedoutAlexander}
	Quantitative analysis indicates that, when the density ratio on both sides of the interface exceeds $\rm10^{4}$, the ARM instability is frozen, previous theoretical models (based on single-layer targets) can accurately describe its evolution.\cite{freezeRM}
	But when this density ratio falls below $\rm10^{4}$, the front surface experiences the ARM instability before the ART instability. 
	Recently, research on the rippled inner-interface-initiated instability in multi-layer targets has commenced. Through simulations, Miller $et$ $al$. \cite{Miller} studied the instability induced by inner perturbations at various depths in multi-layer targets and found that perturbations initially near the ablation front feedout earlier, but their growth rates are basically the same as those initially away from the ablation front.

	Besides, the CH ablator of ICF target is typically doped with certain high-Z element, such as chlorine (Cl), bromine (Br), or silicon (Si). 
	These dopants can suppress the growth rate of the ART instability,\cite{DAF:prl-fujioka, DAF:pop-fujioka, XBBrt, YuanRT} reduce the oscillation period and amplitude of the ARM instability,\cite{XBBRM} help diagnose the target,\cite{DCI-spec} mitigate hot electron preheating,\cite{xuhan2019:ppcf, yang2023:mre} and suppress the laser imprint.\cite{laserimprint} 
	A distinct ablation structure has also been observed in simulations of high-Z doped targets, known as the double ablation front (DAF).\cite{DafTheory1, DafTheory2, DafTheory3} 
	Recent simulations have revealed that high-Z dopants enhance electron, ion, and radiation temperatures, increase the mass ablation rate, and decrease ablation pressure in laser ablation.\cite{Xiong}
	
	As noted previously, inner defects are regarded as detrimental to ICF.
	Studying the evolution of the rippled inner-interface-initiated RT instability can improve our understanding of these defects. However, previous studies have mainly focused on single-layer targets and are unable to accurately describe this instability in ICF multi-layer targets. 
	A theoretical model that can fully describe the evolution of this instability in multi-layer targets, particularly for calculating feedout seeds, is still needed.
	Furthermore, although high-Z dopants have been observed to suppress ART and ARM instabilities, it remains unclear whether they can suppress the rippled inner-interface-initiated instability.
	The aim of this paper is twofold. 
	First, to develop a theoretical model that can quantitatively calculate the feedout seeds and describe the subsequent growth of ARM and ART instabilities, while proposing a potential suppression strategy based on this theoretical model.
	Second, to reveal the mechanisms by which high-Z dopant influence the whole evolution of these instabilities.
	This article is organized as follows: In Sec.\ref{sec2}, we provide a theoretical analysis of the evolution of the rippled inner-interface-initiated ART instability and develop a theoretical model for the feedout seeds, ARM, and ART instabilities. Sec.\ref{sec2.3} presents the physical model used in our simulations. Sec.\ref{sec3} presents the 2D radiation hydrodynamic simulation results for laser ablating high-Z doped targets. The influence of high-Z dopants on the rippled inner-interface-initiated instability is also revealed. In Sec.\ref{sec5}, we further investigate the effects of high-Z dopants on different wavelength rippled inner-interface-initiated instabilities. 
	Finally, Sec.\ref{sec6} provides the conclusions.
	\section{Theory of the Rippled inner-interface-initiated instability}\label{sec2}
	\subsection{Generation of feedout seeds}
	\begin{figure}
		\includegraphics[width=9cm]{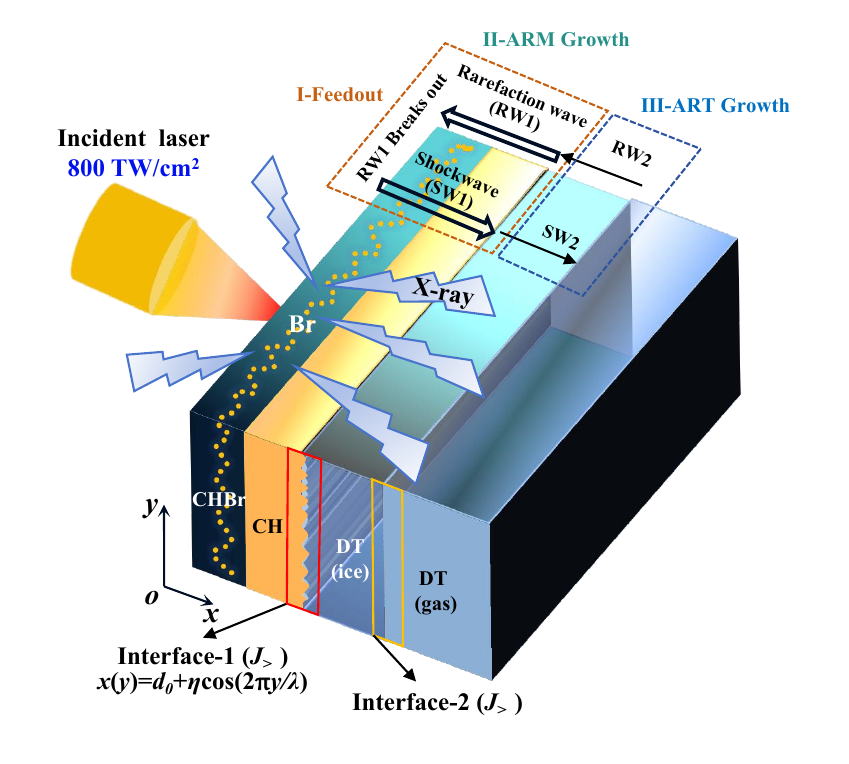}%
		\caption{\label{figure1} (Color online) Schematic of laser-ablating multi-layer high-Z doped target with a rippled CH ablator-DT ice interface, which can be described by [$x(y)=d_0+\eta\cos(2\pi y/\lambda)$]. Interface-1 and Interface-2 represent interfaces between the CH ablator and DT ice, and between the DT ice and DT gas. SW1, SW2 are the incident shock wave and the transmitted shockwave passing through the Interface-1, while RW1 and RW2 denote the rarefaction waves reflected at Interface-1 and Interface-2, respectively.}
	\end{figure}
	An ICF target typically consists of high-Z doped ablator, undoped ablator, DT ice and DT gas, with density decreasing from the outside in, as shown in Fig. \ref{figure1}. It is found that the evolution of the rippled inner-interface-initiated instability consists of three stages: feedout, ARM, and ART instabilities. In the first feedout stage, the seeds are generated on the initially flat front surface, and during the subsequent ARM and ART instabilities stages, these seeds are further amplified. These three stages are also briefly described in Fig. \ref{figure1}. 
	
\begin{figure}
		\includegraphics[width=8cm]{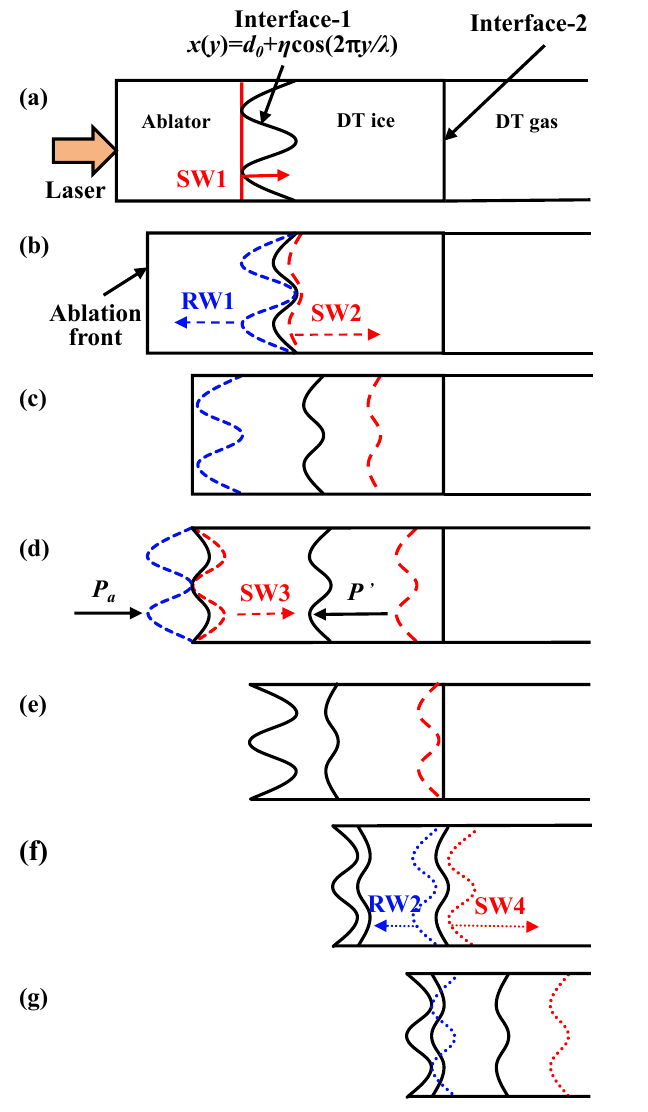}%
		\caption{\label{figure8} (Color online) Schematic diagram of the evolution of rippled inner-interface-initiated instability when SW1 reaches the perturbation valleys (a) and peaks (b), RW1 from valleys breaks out (c), RW1 from peaks breaks out (d), SW2 breaks out (e), RW2 forms (f) and RW2 breaks out (g). SW3 represents the shockwave reflected by RW2 at the ablation front, while SW4 represents the transmitted shock wave of SW2 passing through the Interface-2.}
	\end{figure}

	Figure \ref{figure8} provides a more detailed schematic diagram of the evolution of the rippled inner-interface-initiated instability. The red, blue, and black lines in Fig. \ref{figure8} represent the shockwave, rarefaction wave, and the interface (or ablation front), respectively. 
	Interface-1 and Interface-2 represent interfaces between the CH ablator and DT ice, and between the DT ice and DT gas, respectively. 
	Both of them are $J_>$ type interfaces, where the wave impedance of the transmitted medium is lower than that of the incident medium. For example, the condition $\rho_{\rm CH}\cdot a_{\rm ps(CH)}>\rho_{\rm DT}\cdot a_{\rm ps(DT)}$ is satisfied for Interface-1. 
	SW1, SW2, SW3 and SW4 represent the incident shockwave, the shockwave passing through the Interface-1, the shockwave reflected at the ablation front, and the shockwave passing through the Interface-2.
	RW1, RW2 represent the rarefaction wave reflected at the Interface-1 and Interface-2.

	During the laser ablation, the ablation pressure $P_a$ drives a shockwave propagating inward (SW1 in Fig. \ref{figure1} and Fig. \ref{figure8}(a)). The shockwave speed $D$ and the postshock sound speed $a_{ps}$ are\cite{FeedoutAlexander}
	\begin{equation}\label{eq-1}
		D=\sqrt{\frac{\gamma +1}{2}\frac{P_a}{\rho_0}},
	\end{equation}
	\begin{equation}\label{eq-2}
		a_{ps}=\sqrt{\frac{\gamma P_a}{\rho_{ps}}}=\frac{\sqrt{2\gamma(\gamma-1)}}{\gamma+1}D,
	\end{equation}
	where $\rho_0$, $\rho_{ps}$ and $\gamma$ denote the initial density, the postshock density and the adiabatic exponent of the ablator, respectively. The approximation conclusion for strong shock waves $\rho_{ps}=\rho_{0}(\gamma+1)/(\gamma-1)$ has been used to simplify Eq. (\ref{eq-2}).
	
	When SW1 reaches Interface-1 after a time interval of $t_{sb}=d/D$, where $d$ represents the ablator thickness, a reflected rarefaction wave (RW1) and a transmitted shockwave (SW2) formed at Interface-1, as shown in Figs. \ref{figure8}(a) and \ref{figure8}(b). In the laboratory reference frame, RW1 propagates with a constant velocity $dx/dt$:\cite{FeedoutBetti}
	\begin{equation}
		\frac{dx}{dt}=U_{ps}-a_{ps}
	\end{equation}
	$U_{ps}$ is the postshock speed, which is the speed of the ablation front as well. Therefore, the relative velocity between the ablation front and RW1 is $a_{ps}$. After a time interval of $t_{rb}=d_{ps}/a_{ps}$, where $d_{ps}= d(\gamma-1)/(\gamma+1)$, is the postshock target thickness, RW1 reaches the ablation front and breaks out, simultaneously imparting a lower acceleration $g$ to the ablation front. The time at which RW1 breaks out is:
	\begin{equation}\label{eq-5.1}
		t_{rb} = \frac{d}{D}+\frac{d_{ps}}{a_{ps}} = (1+\sqrt{\frac{\gamma-1}{2\gamma}})\sqrt{\frac{2}{\gamma+1}}\frac{d}{\sqrt{P_a/\rho_0}}
	\end{equation}
	As shown in Fig. \ref{figure8}(a), Interface-1 is rippled in the form $x(y)=d_0+\eta\cos(2\pi y/\lambda)$, where $\lambda$ and $\eta$ are the perturbation wavelength and amplitude, $d_0$ is the initial thickness of the ablator.
	Due to this ripple, the target thickness ($d$) varies between the perturbation valleys ($d_{valley}=d_0-\eta$) and peaks ($d_{peak}=d_0+\eta$), thereby forming a rippled rarefaction wave (RW1).  
	The valleys of Interface-1 reflect the rarefaction wave earlier, thus forming the peaks of RW1. Later, the peaks of Interface-1 reflect the rarefaction wave, forming the valleys of RW1, as blue dotted lines shown in Fig. \ref{figure8}(b). 
	The rippled RW1 causes its valleys and peaks to break out at different times, as shown in Fig. \ref{figure8}(c).
	When the peaks of RW1 break out at the ablation front (Fig. \ref{figure8}(c)), the feedout stage begins, while the breakout of the valleys of RW1 marks its end (Fig. \ref{figure8}(d)).
	Therefore, the feedout duration is $\Delta t = t_{rb,peak}-t_{rb,valley}$, which can be further calculated using Eq. (\ref{eq-5.1}):
	\begin{equation}\label{eq-dt}
		\Delta t =  (1+\sqrt{\frac{\gamma-1}{2\gamma}})\sqrt{\frac{2}{\gamma+1}}\frac{2\eta}{\sqrt{P_a/\rho_0}}
	\end{equation}
	
	As shown in Figs. \ref{figure8}(c) and \ref{figure8}(d), when the peaks of RW1 break out, the ablation front starts to accelerate at this location, while in other locations, it continues to move at the post-shock velocity ($U_{ps}$). Consequently, feedout seeds gradually appear on the initially flat ablation front. 
	Assuming the ablation front at the perturbation valley is uniformly accelerated, then the velocity perturbation and interface perturbation can be expressed as $v_0\approx g\Delta t$, $x_0\approx 1/2 g\Delta t^2$, respectively. The feedout duration $\Delta t$ is given by Eq. (\ref{eq-dt}), and the acceleration $g$ must also be determined to calculate $v_0$ and $x_0$.
	
	For a compressed ablator, the pressure experienced includes the ablation pressure ${P_a}$ and the pressure of SW2 ($P^{\prime}$), as shown in Fig. \ref{figure8}(d). So the acceleration $g$ can be calculated by:
	\begin{equation}\label{eq-g}
		g =  \frac{P_a(1-n)}{\sum_{i} \rho_i d_i}
	\end{equation}
	where $n=P^{'}/P_a$. The denominator $\sum_i\rho_id_i$ represents the total ablator mass, where subscript $i$ denotes different ablator materials. Using Eqs. (\ref{eq-dt})-(\ref{eq-g}), the velocity perturbation ($v_0$) and interface perturbation ($x_0$) can be expressed as:
	\begin{equation}\label{eq-6}
		v_0 = (1+\sqrt{\frac{\gamma-1}{2\gamma}})\sqrt{\frac{2}{\gamma+1}}\frac{2\eta \sqrt{P_a\rho_0}}{\sum_{i} \rho_i d_i}(1-n)
	\end{equation}
	\begin{equation}\label{eq-7}
		x_0 = \frac{(1+\sqrt{\frac{\gamma-1}{2\gamma}})^2}{\gamma+1}\frac{4\eta^2 }{\sum_{i} d_i}(1-n)
	\end{equation}
	
	It is found that the velocity feedout seeds $v_0$ mainly depends on $P_a$ and $\gamma$, while the interface feedout seeds $x_0$ mainly depends on the ablator thickness $\sum_{i} d_i$ and $\gamma$. As $\gamma$ decreases, both $v_0$ and $x_0$ increases. 
	Additionally, both $v_0$ and $x_0$ are related to $P_a$ and $P^{\prime}$. 
	Higher $P_a$ results in more severe feedout seeds as well.
	$P_a$ depends on the incident laser and the ablator material, while $P^{\prime}$ (or $n$) depends on both $P_a$ and the properties of the transmitted material (DT ice). 
	The accurate solution of $n$ requires a method known as impedance matching,\cite{shockwaveTheory} which involves determining the intersection of the isentropic unloading curves of the incident material (CH) and the shock compression curves of the transmitted material (DT ice) in the $P$-$u$ space. 
	The Hugoniot relations are essential for determining both the shock compression and isentropic unloading curves.
	In ICF, the linear Hugoniot relations become invalid due to the extremely high pressures, typically ranging from tens to hundreds of Mbar. 
	Therefore, quadratic or higher-order Hugoniot relations must be considered. However, accurately determining the coefficients for these higher-order relations in such high-pressure conditions is challenging, making it difficult to obtain precise values for ($n$).
	Considering the impedance of DT ice is much smaller than that of CH, we can approximate that $P_a$ is much larger than $P^{'}$. 
	Consequently, the feedout seeds, $v_0$ and $x_0$, can be estimated using Eqs. (\ref{eq-6})-(\ref{eq-7}), with $n$ approximated as zero. 
	Following simulation results indicate that the error introduced by this approximation is within an acceptable range.

	\subsection{Amplification of ARM and ART instabilities}
	
	As shown in Fig. \ref{figure8}(d), the ablation front becomes non-flat after the feedout stage, exhibiting both velocity perturbation ($v_0$) and interface perturbation ($x_0$). 
	The laser continues to incident on the rippled ablation front, and the evolution of the perturbation enters the ARM instability stage, starting from $t$ = $t_{rb, peak}$. 
	The solution of ARM instability has the form of $[C\cos(\omega t)+D\sin(\omega t)]\rm{exp}(-2\it{kv_at})$, where $C$ and $D$ are two constants determined by initial conditions.\cite{ICFYang}
	The initial conditions for the ARM instability are $d\eta(t)/dt(0)$ = $v_0$, $\eta(0)$ = $x_0$, so the time evolution of perturbation can be solved:
	
	\begin{widetext}
		\begin{align}\label{eq-8.1}
			\eta(t) = [x_0\cos(\omega (t-t_{rb,peak}))+
			\frac{v_0+2kv_ax_0}{\omega}\sin(\omega (t-t_{rb,peak}))]e^{-2kv_a(t-t_{rb,peak})}
		\end{align}
		Alternatively, it can be expressed in a form that is more analytically convenient:
		\begin{align}\label{eq-8.2}
			\eta(t) = [x_0\sqrt{1+\frac{4k^2v_a^2}{\omega^2}}\cos(\omega (t-t_{rb,peak})-\phi)+\frac{v_0}{\omega}\sin(\omega (t-t_{rb,peak}))]e^{-2kv_a(t-t_{rb,peak})}
		\end{align}
	\end{widetext}
	where $\omega=kA_T\sqrt{v_av_b}$, is the oscillation frequency of the ARM instability, $A_T$ is the Atwood number, $v_a$ and $v_{b}$ are the ablation velocity and blow off velocity, $\phi$ = $\tan^{-1} (2kv_a/\omega)$. 
	It is important to note that the perturbations on Interface-1 will evolve according to the classical Richtmyer-Meshkov (RM) instability theory, as they are not directly ablated by the laser. 
	These perturbations experience once phase flip, as shown in Figs. \ref{figure8}(e) and \ref{figure8}(f). 
	As the ablation front gradually approaches Interface-1, the ablation effect becomes significant. 
	The perturbations on Interface-1 and ablation front evolve according to the same ARM instability.
	
	The next evolution stage is the ART instability. As shown in Figs. \ref{figure8}(e) and \ref{figure8}(f), when SW2 reaches Interface-2, RW2 is reflected to the ablation front. In fact, the perturbation amplitude of RW2 in Fig. \ref{figure8}(f) is exaggerated and it can be considered flat compared to the ablation front perturbation. Therefore, it can be approximated that the peaks and valleys of RW2 reach the ablation front simultaneously.
	After RW2 reaches the ablation front, the ablator and compressed DT ice move with a high acceleration. This results in the exponential growth of surface perturbations:
	\begin{equation}\label{eq-RM}
		\eta(t) = \eta_{ARM}e^{\gamma^\star [t-(t_{rb,peak}+\delta t)]}
	\end{equation}
	here, $\eta_{ARM}$ is the final perturbation during the ARM instability stage, which is obtained by substituting $t-t_{rb,peak}$ in Eq. (\ref{eq-8.2}) with the ARM instability duration, $\delta t$. After this substitution, the time evolution of the perturbation can be expressed as:
	\begin{widetext}
		\begin{align}\label{eq-9.2}
			\eta(t) = 
			[x_0\sqrt{1+\frac{4k^2v_a^2}{\omega^2}}\cos(\omega \delta t-\phi)+\frac{v_0}{\omega}\sin(\omega \delta t)
			]e^{\gamma^\star[(t-(\delta t+t_{rb,peak})]-2kv_a\delta t}
		\end{align}
	\end{widetext}
	$\gamma^\star$ in the above formula is the perturbation growth factor, which can be described by various models. \cite{Takabe,GoncharovSelfConsistentLargeFroudRT,GoncharovSelfConsistentSmallFroudRT}
	
	To summarize, a theoretical model that divide the whole evolution into three stages (feedout, ARM and ART instabilities) has been developed. 
	The feedout seeds ($v_0$ and $x_0$) can be estimated by Eqs.(\ref{eq-6})-(\ref{eq-7}), while the subsequent ARM and ART instabilities can be described by Eqs.(\ref{eq-8.2}) and (\ref{eq-9.2}). 
	Theory suggests that the feedout seeds mainly depend on the early ablation pressure and the adiabatic exponent, while the growth rates of subsequent ARM and ART instabilities mainly depend on the ablation velocity. 
	An ablator with lower early ablation pressure, a high adiabatic exponent, and a high ablation velocity can reduce both the feedout seeds and the growth rate of ARM/ART instabilities, thereby significantly suppressing this instability. 
	This may be a potential strategy to suppress the rippled inner-interface-initiated instability, but further research is still needed to confirm its validity.
	Additionally, it is important to note that our theoretical model about the feedout seeds is only valid when the long-wavelength condition ($kd_{ps}<1$) is satisfied. 
	For shorter wavelength perturbations ($kd_{ps}>$1), two-dimensional effects, such as the lateral mass flow, must be considered.\cite{FeedoutAlexander} In this case, the feedout seeds will be lower than our theoretical predictions due to the lateral mass flow, and further research is still needed.
	\section{Physical model and simulation settings}\label{sec2.3}
	Based on the Euler code FLASH,\cite{FLASH} 2D radiation-hydrodynamic simulations are performed to validate the theory of the rippled inner-interface-initiated ART instability that we proposed in last section. 
	This section mainly outlines the basic physics models, target and laser parameters, as well as the equations of state (EOS) and opacity, that we used in simulations.
	
	Simulation of the rippled inner-interface-initiated ART instability in laser-ablating high-Z doped targets involves laser ablation and radiation transport, thus requiring the radiation hydrodynamic simulations. 
	Radiation hydrodynamic simulations are commonly used to study fluid dynamics, particularly in cases involving extreme interactions between radiation and matter, such as hydrodynamic instabilities in ICF \cite{ChenZeHaoNF,XuBiHaoHPL} and supernova explosions.\cite{WangReview} 
	Radiation hydrodynamic simulations solve the coupled equation system of hydrodynamic and radiation transport equations. 
	The system of hydrodynamic equations consists of three conservation laws: mass conservation, momentum conservation, and energy conservation. The mass conservation and momentum conservation equations are:
	\begin{equation}\label{mass}
		\frac{\partial \rho}{\partial t} +\nabla\cdot(\rho \mathbf{v})=0,
	\end{equation}
	\begin{equation}\label{mom}
		\frac{\partial (\rho\mathbf{v })}{\partial t} +\nabla(P_\mathrm{i} + P_\mathrm{e} +P_\mathrm{r} ) +\nabla\cdot(\rho \mathbf{vv} )=0,
	\end{equation}
	where $\rho$ is the fluid density, $\mathbf{v}$ is the fluid velocity, and $P_i$, $P_e$, $P_r$ are the pressure of ions, electrons and radiation. Since the electrons, ions, and radiation are not in thermal equilibrium, the energy equation is divided into three separate equations:
	\begin{align}\label{subeq:3}
		\frac{\partial(\rho \epsilon_\mathrm e)}{\partial t}+\nabla\cdot(\rho \epsilon_\mathrm e \mathbf{v})+P_\mathrm{e}\nabla\cdot \mathbf{v}=\rho\frac{c_{v,e}}{\tau_{ei}}(T_\mathrm{i}-T_{\mathrm e})-\nabla\cdot \mathbf {q_e} \nonumber \\
		+Q_{\mathrm{abs}}-Q_{\mathrm {emis}}+Q_{\mathrm{las}}
	\end{align}
	\begin{equation}\label{subeq:4}
		\frac{\partial(\rho \epsilon_\mathrm i)}{\partial t}+\nabla\cdot(\rho \epsilon_\mathrm i \mathbf{v})+P_\mathrm{i}\nabla\cdot \mathbf{v}=\rho\frac{c_{v,e}}{\tau_{ei}}(T_\mathrm{e}-T_{\mathrm i})
	\end{equation}
	\begin{equation}\label{subeq:5}
		\frac{\partial(\rho \epsilon_\mathrm r)}{\partial t}+\nabla\cdot(\rho \epsilon_\mathrm r \mathbf{v})+P_\mathrm{r}\nabla\cdot \mathbf{v}=\nabla\cdot \mathbf {q_r}-Q_{\mathrm{abs}}+Q_{\mathrm {emis}}
	\end{equation}
	where $\epsilon$ is the specific internal energy, and the subscripts $i,e,r$ represent the variables corresponding to ions, electrons and radiation, respectively. The terms on the right-hand side of Eqs. (\ref{subeq:3})-(\ref{subeq:5}) represent different energy absorption and loss mechanisms. $c_{v,e}/\tau_{ei}(T_\mathrm{i}-T_{\mathrm e})$ in Eqs. (\ref{subeq:3})-(\ref{subeq:4}) is the electron-ion interaction term, while $c_{v,e}$ is the electron specific capacity, $\tau_{ei}$ is the relaxation time. 
	$Q_{\mathrm{las}}$ represents the energy absorbed from the laser, determined by the reverse bremsstrahlung absorption power. The laser tracing module is also used.  $Q_{\mathrm{emis}}$ and $Q_{\mathrm{abs}}$ are the radiation emission, and the radiation absorption, respectively. 
	The electrons or radiation can also lose energy by the electron heat flux $-\nabla\cdot \mathbf {q_e}$ and radiation heat flux $-\nabla\cdot \mathbf {q_r}$.
	
	Radiation terms $Q_{\mathrm{abs}}$, $Q_{\mathrm {emis}}$ describe the energy exchange between the radiation field and electrons, which requires solving the radiation transport equation. 
	For convenience, we adopt the multi-group diffusion (MGD) method, and $Q_{\mathrm{abs}}$ and $Q_{\mathrm {emis}}$ for group g are given by: 
	\begin{equation}
		Q_{\rm{abs,g}}=cu_{\rm{r,g}}\sigma_{\rm{a,g}} 
	\end{equation}
	\begin{equation}\label{last}
		Q_{\rm{emis,g}}=c\sigma_{\rm{e,g}}aT_e^4\frac{15}{\pi^4}[P(x_{\rm{g+1}})-P(x_{\rm{g}})]
	\end{equation}
	Subscript g and g+1 represent the variables of the corresponding groups. The quantities $u_{\rm{r,g}}$, $\sigma_{\rm{a,g}}$ and $\sigma_{\rm{e,g}}$ denote the radiation energy density, absorption opacity, and emission opacity, respectively. Here, $a=4\sigma/{c}$ is the radiation constant, where $\sigma$ is the Stefan-Boltzmann constant, and $c$ is the speed of light. The variable $x_{\rm{g}}$ is defined as $(h\nu)_{\rm{g}}/k_BT_e$, where $h$ is the Planck constant and $k_B$ is the Boltzmann constant. The function $P(x)$ is defined as $P(x)=\int_{0}^{x}(x')^3/(\rm{exp}(\it{x'})-\rm{1})dx'$.
	
	Additionally, the EOS and opacity of materials are required to solve Eqs. (\ref{mass})-(\ref{last}). 
	The EOS incorporates several effects, including the electron degeneracy, the Debye-Huckel correction, the ionization and excitation, and the Quotidian Equation of State (QEOS) model.\cite{More} These considerations enable an accurate description of the EOS at densities near or below the solid density.
	The opacity values are determined from the parameters in ATBASE database.\cite{macfarlane2006:joqsrt,wangping1991:phd} Both bound-bound, bound-free, and free-free contributions to the opacity are included. 
	
	A typical laser direct-driven ICF target configuration has been shown in Fig. \ref{figure1}. In this article, we select $\rm C_{50}H_{47}Br_3$ with a 3\% bromine (Br) doping ratio as the high-Z dopant of the CH ablator. Detailed parameters are listed in Table \ref{table1.1}. For comparison, the undoped target consists of three layers: the CH ablator, DT ice, and DT gas. We increase the thickness of the ablator in the undoped target ($8.16$ $\mu\rm m$) to maintain the same ablator mass as the doped targets. 
	The intensity of the square wave laser is 800 $\rm TW\cdot cm^{-2}$, uniformly distributed across the laser focal spot. 
	Above target and laser parameters are referenced from the relevant experiment setup on OMEGA. \cite{OmegaExperiment01,OmegaExperiment02}
	For convenience, a planar target model is adopted instead of a spherical one, as convergence effects are negligible in the early stages of laser direct-driven ICF.
	
	\begin{table}
		\caption{\label{table1.1}Parameters of the high-Z doped target}
		\begin{ruledtabular}
			\begin{tabular}{ccccc}
				Material of & $A$&$Z$&$\rho$&$d$ \\
				each layer &&&($\rm g\cdot \rm{cm^{-3}})$&$\rm (\mu \rm{m})$\\
				\hline
				$\rm C_{50}\rm H_{47}\rm Br_3$&$8.87$&$4.52$&$1.26$&$5.20$\\
				$\rm C_{16}\rm H_{16}$&$6.50$&$3.50$&$1.1$&$2.20$\\
				$\rm DT$ $\rm ice$&$1.5$&$1$&$0.22$&$47.50$\\
				$\rm DT$ $\rm gas$&$1.5$&$1$&$\rm 3\times 10^{-4}$&$50.00$
			\end{tabular}
		\end{ruledtabular}
	\end{table}

	\section{Simulation results of The rippled inner-interface-initiated instability and the effect of high-Z dopant on it}\label{sec3}
	\begin{figure*}
		\includegraphics[width=18.5cm]{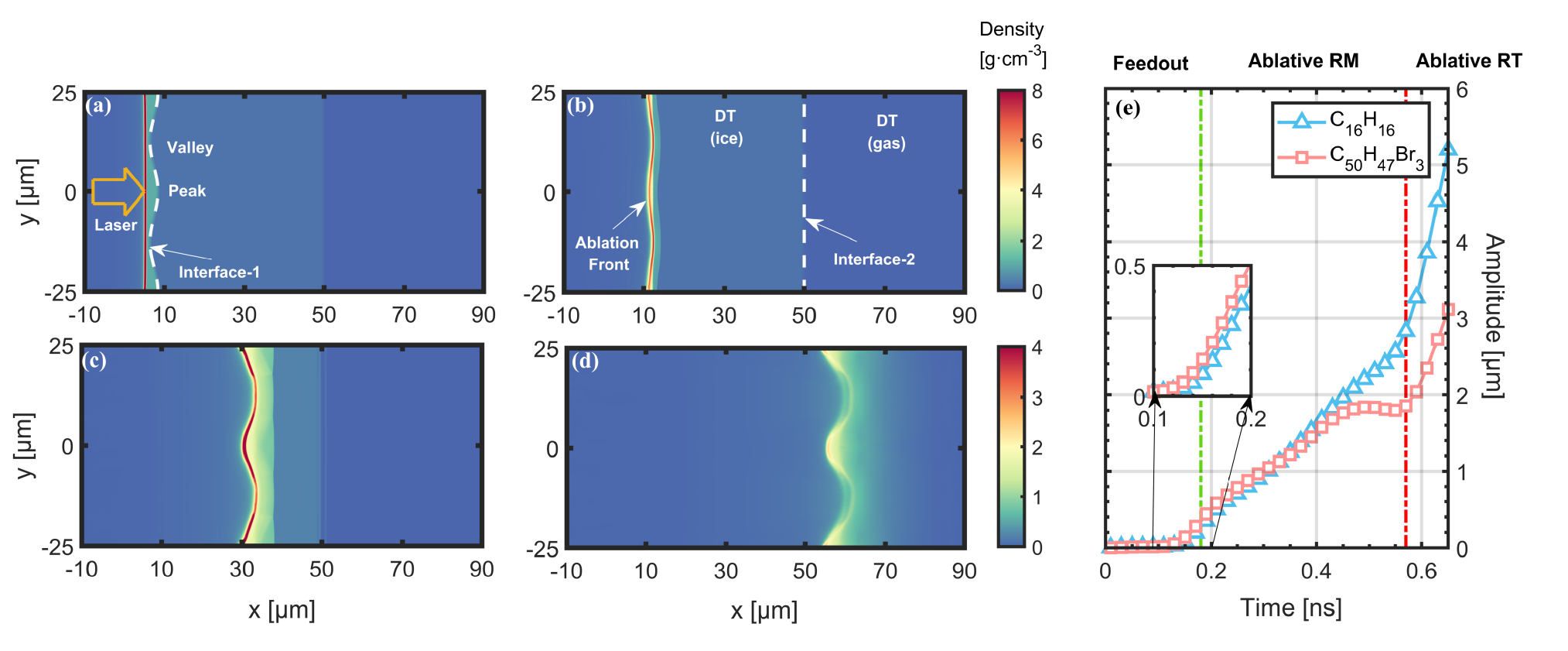}%
		\caption{\label{figure2.1} (Color online) 2D radiation hydrodynamic simulation results of density in high-Z doped targets at 0.10 ns (a), 0.20 ns (b), 0.40 ns (c) and 0.60 ns (d). And the time evolution curves of fundamental mode perturbations at the ablation front in high-Z doped and undoped  targets (e). The white dotted lines in (a) and (b) represent the Interface-1 (interface between CH and DT ice) and Interface-2 (interface between DT ice and DT gas). The green and red vertical dotted lines represent the boundary between the feedout stage and ARM instability, as well as the boundary between the ARM and ART instability, respectively.}
	\end{figure*}
	In this section, the evolution of the rippled inner-interface-initiated ART instability is simulated using 2D radiation hydrodynamic methods, as outlined in Sec.\ref{sec2.3}. 
	Simulation results are in good agreement with the predictions of theoretical model derived in Sec.\ref{sec2}. 
	The effect of high-Z dopant on this process, including the feedout seeds, as well as the subsequent ARM and ART growth rates, are also revealed.

	\subsection{Evolution of the rippled inner-interface-initiated instability in high-Z doped targets}\label{sec4.1}
	
	Figures \ref{figure2.1}(a) and \ref{figure2.1}(d) show the density distributions of the rippled inner-interface-initiated ART instability at different times: 0.10, 0.20, 0.30, and 0.60 ns, corresponding to stages before the feedout, after the feedout, during the ARM instability, and the ART instability. The perturbation has a wavelength of 25 $\rm \mu \rm m$ and an amplitude of 1 $\rm \mu \rm m$. White dotted lines in Figs. \ref{figure2.1}(a) and \ref{figure2.1}(b) denote the Interface-1 (rippled) and Interface-2 (flat), respectively. 
	
	In Fig. \ref{figure2.1}(a), SW1 has not yet reached Interface-1, so RW1 has not broken out, and the ablation front remains flat with no perturbation occurring. 
	In Fig. \ref{figure2.1}(b), RW1 breaks out, and the ablation front is no longer flat, leading to the formation of feedout seeds, which can be described by Eqs. (\ref{eq-6})-(\ref{eq-7}). This corresponds to cases in Figs. \ref{figure8}(c) and \ref{figure8}(d).
	In Fig. \ref{figure2.1}(c), the laser continues to ablate the rippled ablation front. 
	Under the influence of dynamic overpressure, laser ablation, and vorticity convection, the perturbation on the ablation front develops as the ARM instability. This corresponds to cases in Figs. \ref{figure8}(e)-\ref{figure8}(f), which can be described by Eq. (\ref{eq-8.2}).
	When RW2 reaches the ablation front, the perturbation on the ablation front enters the ART instability stage, as depicted in Fig. \ref{figure2.1}(d). During this stage, the amplitude of the instability grows exponentially, as shown in Fig. \ref{figure8}(g), which can be described by Eq. (\ref{eq-9.2}).
	The evolution of the ART instability initiated by the rippled inner-interface in undoped targets is similar to that in the high-Z doped targets (not shown for brevity).
	
	Fourier decomposition is performed on the ablation front, which is selected as the contour line corresponding to half of the peak density.\cite{ChenPRL} Then the amplitude of the fundamental component is the perturbation amplitude. 
	Subsequently, the evolution of the perturbation amplitude is determined, as shown in Fig. \ref{figure2.1}(e). 
	The red squares and blue triangles represent the evolution in high-Z doped and undoped targets. 
	From Fig. \ref{figure2.1}(e), it is seen that the perturbation evolution experiences three distinct stages in both undoped and high-Z doped targets: feedout, ARM, and ART instabilities. 
	The green and red dotted vertical lines mark the boundaries between the feedout stage and ARM instability, as well as between the ARM and ART instability. 
	Above results are basically consistent with the theoretical model developed in Sec.\ref{sec2}. 
	
	A comparison of the two curves in Fig. \ref{figure2.1}(e) reveals that, during the feedout stage and the early stage of the ARM instability, the high-Z dopant enhances the perturbation. 
	This suggests that the high-Z dopant increases the feedout seeds. 
	However, during the middle of the ARM instability stage and throughout the ART instability stage, the perturbation in the high-Z doped targets is lower than that in the undoped targets. 
	This leads to an intersection of the perturbation curves for doped and undoped targets during the ARM instability stage.
	Before this intersection, the perturbation in the doped targets is larger than that in the undoped target. 
	But after this intersection, the perturbation in the doped targets becomes smaller than that in the undoped target.
	Additionally, during the ARM instability stage, the perturbation in high-Z doped targets exhibits oscillations (without phase reversal), which are not observed in undoped targets. 
	A more detailed elucidation of this behavior will be provided in the following sections.
	
	\subsection{The influence of high-Z dopant on the feedout seeds}
	\begin{figure*}
		\includegraphics[width=18cm]{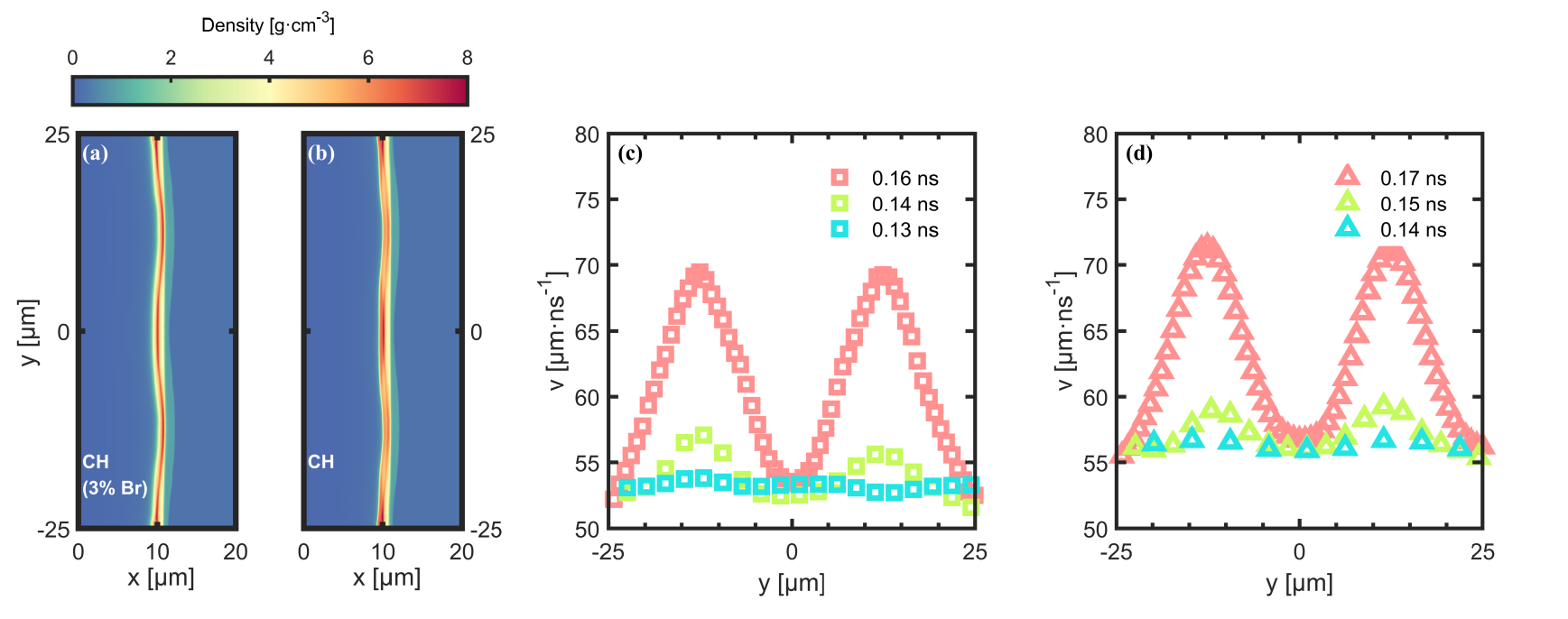}%
		\caption{\label{figure3.1} (Color online) 2D density distributions of high-Z dopant (a) and undoped targets (b) after the feedout (0.18 ns). And the velocity distributions of the high-Z dopant targets (c), as well as undoped targets (d) at different times during the feedout.}
	\end{figure*}
	
	During the feedout stage, perturbations gradually appear on the ablation front. 
	This procedure can be found clearly from the velocity distributions in Figs. \ref{figure3.1}(c)-(d).
	
	Figures \ref{figure3.1}(c) and \ref{figure3.1}(d) are the velocity distributions of the ablation front at different times during the feedout stage for the high-Z doped and undoped targets, respectively. 
	In Fig. \ref{figure3.1}(c), the blue, green, and red squares correspond to the velocity distributions at 0.13, 0.14, and 0.16 ns, respectively, while in Fig. \ref{figure3.1}(d), they represent the results at 0.14, 0.15, and 0.17 ns, respectively.
	
	As blue squares shown in Fig. \ref{figure3.1}(c), the high-Z doped targets do not feedout at 0.13 ns, exhibiting a nearly flat velocity distribution. 
	Starting from 0.14 ns, the perturbations gradually feedout, and the velocity distribution becomes slightly uneven, as shown by green squares in Fig. \ref{figure3.1}(c). It is also observed that the feedout begins from the valleys ($y$ = $\pm$12.5 $\rm{\mu m}$), due to their thinner thickness. 
	At 0.16 ns, the feedout stage in the high-Z doped targets ends, and the velocity perturbation reaches its maximum, as shown by  red squares in Fig. \ref{figure3.1}(c). Meanwhile, the perturbation peaks ($y$ = 0 and $\pm$25 $\rm{\mu m}$) are about to accelerate.  Therefore, the feedout for the high-Z doped targets starts from 0.13 ns and ends at 0.17 ns.
	As shown in Fig. \ref{figure3.1}(d), the evolution of the velocity distributions in the undoped target is similar to that in the high-Z doped targets, with only slight differences observed. Specifically, feedout occurs slightly later in the undoped target, and the velocity feedout seeds are lower compared to those in the high-Z doped targets.
	
	Figures \ref{figure3.1}(a) and \ref{figure3.1}(b) show the density distributions of the high-Z doped and undoped targets at 0.18 ns, corresponding to the end of feedout. 
	It can be observed that the ablation front of the high-Z doped target is more uneven than the undoped target. This suggests that the high-Z dopant may increase the feedout seeds.
	
	Table \ref{table1.3} quantitatively presents the simulation results for the velocity feedout seeds ($\it{v_{\rm{1}}}$) and the interface feedout seeds ($\it{x_{\rm{1}}}$) of $\rm{C_{50}H_{47}Br_3}$ and $\rm{CH}$ targets at the end of feedout. 
	It is found that the velocity seeds $v_1$ in the high-Z doped targets increased by 10.65\%, while the interface seeds $x_1$ increased by 32.88\%. This indicates that the high-Z dopant enhances both the velocity and interface feedout seeds. 
	Table \ref{table1.3} also lists the theoretical results of the velocity and interface feedout seeds ($v_0$ and $x_0$), which is calculated by Eq. (\ref{eq-6}) and Eq. (\ref{eq-7}). 
	The agreement between the theoretical and simulated results confirm the validity of our theory. 
	It is noteworthy that the simulated results ($v_1$ and $x_1$) are slightly lower than the theoretical results ($v_0$ and $x_0$). 
	This may be attributed to the approximation that the postshock pressure $P^{\prime}$ (or $n$) in Eq. (\ref{eq-6}) and Eq. (\ref{eq-7}) is assumed to be zero. 
	The value of $n$ can be calculated precisely, and the accuracy of the theoretical calculations can be further improved, providing that accurate Hugoniot relationships for $\rm{C_{16}H_{16}}$ and DT ice, within a pressure range of tens to hundreds of Mbar, can be determined.
	
	\begin{table*}
		\caption{\label{table1.3} Simulation results of the velocity and interface perturbation seeds ($v_1$ and $x_1$) at the end of feedout stage, as well as the theoretical predictions ($v_0$ and $x_0$), and relevant parameters used to determine $v_0$ and $x_0$.}
		\begin{ruledtabular}
			\begin{tabular}{cccccccccc}
				Target &$\it \gamma $&$\it P_a$&$\it \rho_0$&$\it d_0$ & $\it v_0$ &$\it x_0 $&$\it v_1 $& $\it x_1$\\
				type&& $\rm (Mbar) $&$\rm(g\cdot \rm cm^{-3})$ &$\rm (\mu \rm m)$ &$\rm (\mu \rm m\cdot ns^{-1})$&$\rm (\mu \rm m)$&$\rm (\mu \rm m\cdot ns^{-1})$&$\rm (\mu \rm m)$\\
				\hline
				$\rm C_{50} \rm H_{47}Br_3$&$1.6$&$38.31$&$1.21$&$7.40$&$19.09$&$0.4269$&$16.11$&$0.3605$\\
				$\rm C_{16} \rm H_{16}$    &$2.3$&$31.15$&$1.10$&$8.16$&$15.56$&$0.3485$&$14.56$&$0.2713$\\
			\end{tabular}
		\end{ruledtabular}
	\end{table*}
	
	Next, we apply the obtained theoretical model to elucidate the influence mechanism of high-Z dopant on the feedout seeds.
	Eq. (\ref{eq-6}) indicates that for targets with the same ablator mass ($\sum_{i} \rho_i d_i$), the velocity feedout seeds primarily depend on the ablation pressure $P_a$ and a parameter expression contained $\gamma$. 
	Thus, the influence mechanism of high-Z dopant on the velocity feedout seeds can be attributed to the ablation pressure and $\gamma$ during the feedout stage.
	
	The first row of Table \ref{table1.3} shows the adiabatic exponent $\gamma$ in the high-Z doped and undoped ablators, as obtained from the simulations. $\gamma$ in the doped target is lower than that in the undoped target. 
	Due to the expression $(1+\sqrt{(\gamma-1)/2\gamma})\sqrt{2/(\gamma+1)}$ in Eq. (\ref{eq-6}) decreases as $\gamma$ increases, the decreased $\gamma$ in the high-Z doped targets lead to an increased $v_0$. 
	The decreased $\gamma$ is attributed to the higher ionization level ($Z$) of the doped ablator. 
	Fundamental thermodynamic relationships provide $\gamma=C_p/C_v$, where $C_p$ and $C_v$ represent the isobaric and isochoric molar heat capacities, respectively. Using the Meyer formula, $C_p=C_v+R$, so $\gamma$ is further simplified to $\gamma=1+R/C_v$, where $R$ $\approx $ 8.314 $\rm J\cdot mol^{-1}\cdot K^{-1}$, is the universal gas constant. 
	Therefore, $\gamma$ is inversely proportional to $C_v$. 
	For the high-Z doped ablator $\rm C_{50}H_{47}Br_3$, due to the higher ionization level, more electrons are ionized. 
	Consequently, $\rm C_{50}H_{47}Br_3$ exhibits a higher $C_v$, resulting in a decreased $\gamma$.
	
	\begin{figure}
		\includegraphics[width=7.5cm]{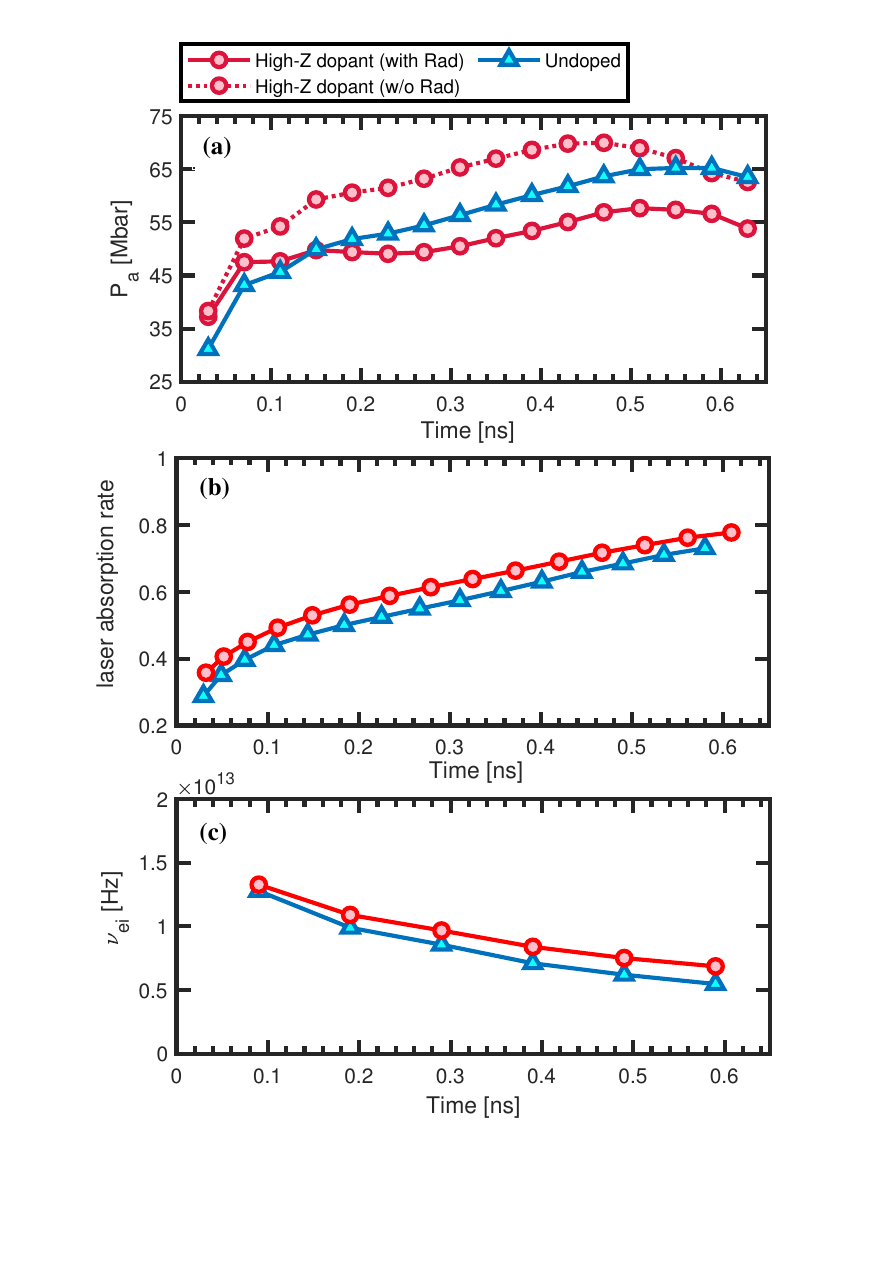}%
		\caption{\label{figure4.1} (Color online) Time evolution of the ablation pressure (a), laser absorption rate (b) and electron-ion collision frequency (c).}
	\end{figure}

	Figure \ref{figure4.1}(a) shows the time evolution of $P_a$. The blue triangles represent the results for the undoped targets, while the red circles represent the results for the high-Z doped targets with radiation (solid line) or without radiation (dotted line).
	As Fig. \ref{figure4.1}(a) shown, during the feedout stage, $P_a$ in the high-Z doped targets is higher than that in the undoped targets. 
	Average $P_a$ in the high-Z doped and undoped targets are calculated in Table \ref{table1.3}. 
	It is found that high-Z dopant increases $P_a$ by 22.99\% during the feedout stage. 
	Therefore, the increased early $P_a$ in the high-Z doped targets enhances the acceleration of the ablation front after RW1 breaks out, leading to a higher velocity feedout seeds.
	
	The mechanism that the early $P_a$ is increased by high-Z dopants can be attributed to higher laser absorption and the density at critical surface $\rho_c$. 
	Comparing the solid and dotted lines with red circles in Fig. \ref{figure4.1}(a), the early evolution of $P_a$ remains almost the same, regardless of whether the radiation module is used. 
	This indicates that the radiation effect is relatively weak during the feedout. 
	Consequently, the steady-state ablation theory, which ignores the radiation, can be used to describe the early $P_a$:\cite{Xiong}
	\begin{equation}\label{eq-Pa}
		P_a=\frac{1}{2^{1/3}}m_p^{1/3}(\frac{n_{c0}}{\lambda_L^{2}})^{1/3}(\frac{A}{Z})^{1/3}I_{abs}^{2/3} \sim \rho_c^{1/3}I_{abs}^{2/3}
	\end{equation}
	here, $I_{abs}$ represent the absorbed laser intensity. $m_p$ is the proton mass, and $n_{c0}/\lambda_L^{2}$ is the electron number density at the critical surface, with $\lambda_L$ = 0.351 $\mu m$ is the wavelength of the incident laser. Eq. (\ref{eq-Pa}) shows that the early ablation pressure primarily depends on $I_{abs}$ and $\rho_c$.
	
	Then consider the effect of high-Z dopant on $I_{abs}$ and $\rho_c$. 
	The laser transfers its energy to the target primarily through the inverse bremsstrahlung absorption with electrons, which depends on the electron-ion collision frequency $\nu_{ei}$:
	\begin{equation}\label{eq-Iabs}
		\nu_{ei}=\frac{4}{3}\left (   \frac{2\pi}{m_e}\right )^{1/2}\frac{n_eZe^4\rm{ln}(\Lambda)}{k_BT_e^{3/2}}
	\end{equation}
	where $m_\mathrm{e}$ is the electron mass, $n_\mathrm{e}$ and $e$ represent the electron number density and the electron charge, $k_B$ is the Boltzmann constant, $\rm ln(\Lambda)$ = $\rm ln[3/(2\it Ze^{\rm 3})\sqrt{k_B^{\rm3}T_\mathrm{e}^{\rm3}/(\pi n_\mathrm{e})}]$, is the coulomb logarithm.\cite{nrlFormular}
	Using Eq. (\ref{eq-Iabs}), $\nu_{ei}$ for high-Z doped and CH targets are calculated, as indicated by the red circles and blue triangles in Fig. \ref{figure4.1}(c). 
	After doping Br atoms, the average ionization level $Z$ increases, leading to $\nu_{ei}$ increases. Thereby, the laser absorption rate ($I_{abs}/I_{l}$) is increased, as shown in Fig. \ref{figure4.1}(b). 
	
	$\rho_c$ is calculated by $n_c/ N_A\cdot\it A/Z$, where $ N_A$ = 6.02$\times$$\rm10^{23}$ $\rm mol ^{-1}$ is the Avogadro constant, and $n_{c}=1.1 \times 10^{21}/\lambda_L^2$ $\rm{cm}^{-3}$ is the electron number density at the critical surface. Here, $A$ represent the average atomic mass. 
	For a specific wavelength of incident laser, $\rho_c$ depends solely on $A/Z$. 
	The values of $A$ and $Z$ for $\rm{C_{50}H_{47}Br_3}$ and $\rm{C_{16}H_{16}}$ are listed in Table \ref{table1.1}. It is found that the $A/Z$ value in the high-Z doped targets is 1.9624, which is 5.95\% higher than that in the undoped targets (1.8571), resulting in an increase in $\rho_c$. 
	Consequently, both $\rho_c$ and $I_{abs}$ increase after doping Br atoms. 
	Then the early $P_a$ is increased.
	
	Eq. (\ref{eq-7}) indicates that the interface feedout seeds $x_0$ mainly depends on the ablator thickness $d_0$ and $\gamma$. 
	As $d_0$ decreases or $\gamma$ decreases, the interface feedout seeds is increased. 
	The mechanism how high-Z dopant decreases $\gamma$ has been elucidated previously. 
	The effect of high-Z dopant on the thickness is relative straightforward.
	Generally, the ablator density increases with high-Z doping. 
	For example, the density of undoped $\rm{C_{16}H_{16}}$ is 1.1 $\rm{g\cdot cm^{-3}}$, whereas the density of $\rm{C_{50}H_{47}Br_3}$ is 1.26 $\rm{g\cdot cm^{-3}}$. Therefore, for targets with the same ablator mass, the ablator in the high-Z doped targets is thinner than that in the undoped targets. 
	This results in an earlier feedout start and a longer feedout duration $\Delta t$, ultimately resulting in a larger interface feedout seed. 
	
	\begin{figure*}\suppressfloats\centering
		\includegraphics[width=14cm]{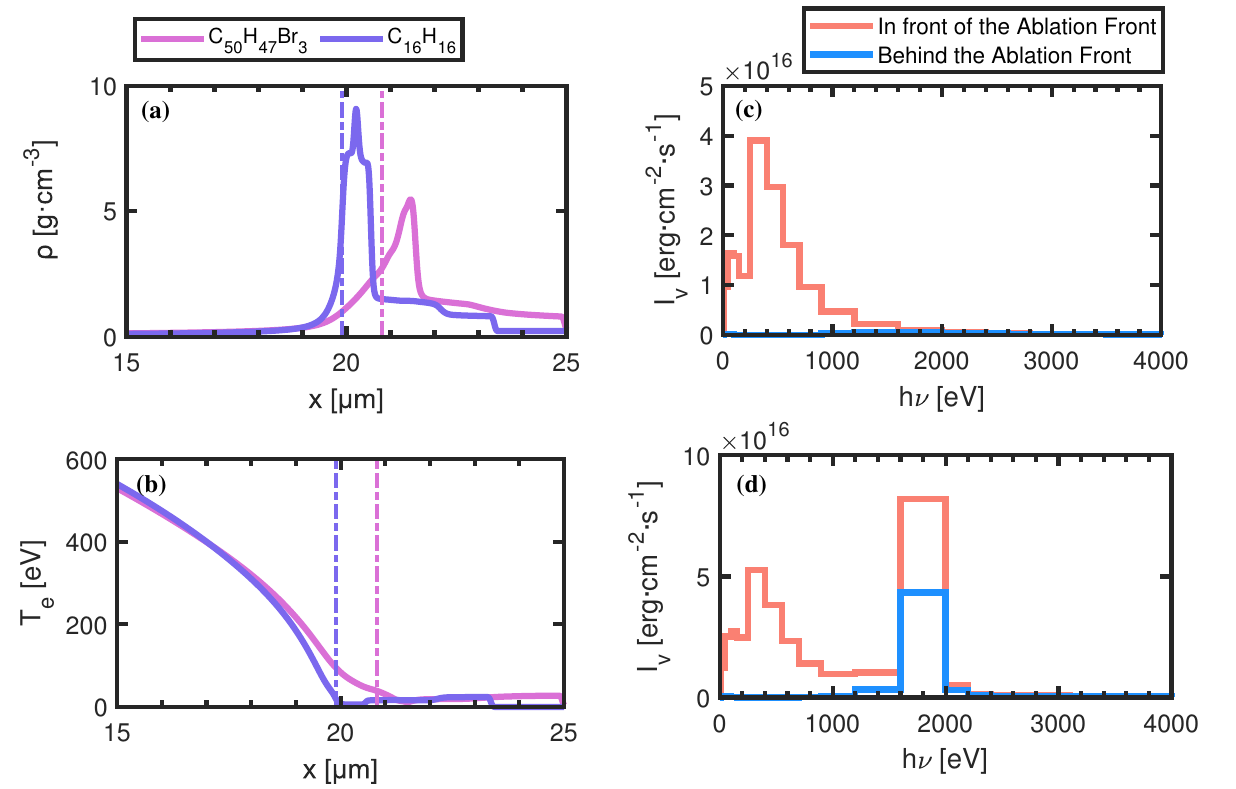}%
		\caption{\label{figure5} (Color online) Distribution of density (a) and electron temperature (b) at 0.30 ns. And radiation spectra in the undoped (c) and high-Z doped targets (d). The vertical dotted lines in (a) and (b) represent the locations of their ablation fronts.}
	\end{figure*}
	\subsection{The influence of high-Z dopant on the perturbation growth rate}
	The perturbation growth rate, $\gamma^\star$, during the ART instability stage can be obtained through exponential fitting of the perturbation evolution curves, as shown in Table \ref{table4}. It is observed that the ART instability growth rate is reduced by 16.82\% due to the high-Z dopant.
	\begin{table}
		\caption{\label{table4} Time average parameters calculated by Betti fitting method and the ART instability growth factor}
		\begin{ruledtabular}
			\begin{tabular}{cccc}
				Target&$\it v_a$ &$\it L_m$ &$ \it \gamma^\star $\\
				type&($\rm \mu \rm m\cdot\rm ns^{-1})$&($\rm \mu \rm m$)&($ns^{-1}$)\\
				\hline
				Undoped&$2.21$&$0.51$&$7.73$\\
				High-Z doped&$5.73$& $0.80$ &$6.43$\\
			\end{tabular}
		\end{ruledtabular}
	\end{table}
	
	For perturbation with a wavelength of 25 $\mu\rm m$, ablation strongly suppresses the ARM and ART instability growth rates in the high-Z doped targets.\cite{GoncharovRM01,GoncharovRM02} During the ARM instability, ablation exhibits an exponential decay coefficient of $-2kv_a$ in Eq. (\ref{eq-8.1}), while during the ART instability stage, the ablation exhibits a linear decrease of $-kv_a$, according to the Takabe-Bodner-Lindl formula, $\gamma^\star=\alpha\sqrt{kg/(1+kL_m)}-\beta kv_a$.
	Using the Betti fitting procedure,\cite{BettiFitting} the ablation velocity $v_a$ and the minimum scaled density length $L_m$ are obtained, as listed in Table \ref{table4}. 
	$L_m$ is calculated by $L_0\times(\nu+1)^{(\nu+1)}/(\nu^\nu)$, where $L_0$ is the characteristic thickness of the ablation front and $\nu$ is the power index of the thermal conductivity, both of which are obtained through the Betti fitting procedure.
	From Table \ref{table4}, $v_a$ in the high-Z doped targets is significantly higher than that in the undoped targets. This increase is due to the enhanced radiation, or larger radiation opacity $\sigma_{e,g}$ in the high-Z dopant.
	
	Figures \ref{figure5}(c) and \ref{figure5}(d) show the radiation spectra of the undoped and high-Z doped targets.
	The red and blue lines represent the spectra measured in front of and behind the ablation front, respectively. The spectrum in front of the ablation front was measured at a distance of 2.5 $\mu\rm m$ away from the ablation front, while the spectrum behind the ablation front was measured at a distance of 2 $\mu\rm m$ away from the ablation front.
	
	From Figs. \ref{figure5}(c) and \ref{figure5}(d), it is found that the radiation emitted from the high-Z doped targets is predominantly the hard X-rays, whereas the radiation from the undoped targets mainly consists of the soft X-rays.
	The energy of the hard X-rays in the Br-doped targets ranges from 1600 to 2000 eV, corresponding to the L-band radiation of Br atoms.
	Additionally, soft X-rays have a shorter mean free path than hard X-rays, thus it is more difficult for them to pass through the ablation front. 
	Therefore, the intensity of soft X-rays decreases sharply as they pass through the ablation front, approaching zero, as shown by the low-energy portion of the blue lines in Figs. \ref{figure5}(c) and \ref{figure5}(d).
	In contrast, the hard X-rays can easily pass through the ablation front and reach the cold target, with only a slight decrease in intensity.
	
	Figures \ref{figure5}(a) and \ref{figure5}(b) show the distributions of $\rho$ and $T_e$ at 0.30 ns for the high-Z doped and undoped targets, respectively. 
	The purple lines represent the results for the undoped targets, while the magenta lines represent the results for the high-Z doped targets. 
	Two vertical dashed lines represent the position of the ablation front. 
	As shown in Fig. \ref{figure5}(b), high-energy X-rays induce severe preheating, which increases $T_e$ inside the target and decreases the $T_e$ gradient near the ablation front. 
	Consequently, the peak compression density decreases, as shown in Fig. \ref{figure5}(a). 
	This results in a reduction of the density gradient, which in turn leads to an increase in $L_{m}$, as shown in Table \ref{table4}. This increase in $L_{m}$ also contributes to a decrease in the perturbation growth rate.

	\begin{figure*}\suppressfloats\centering
		\includegraphics[width=12.5cm]{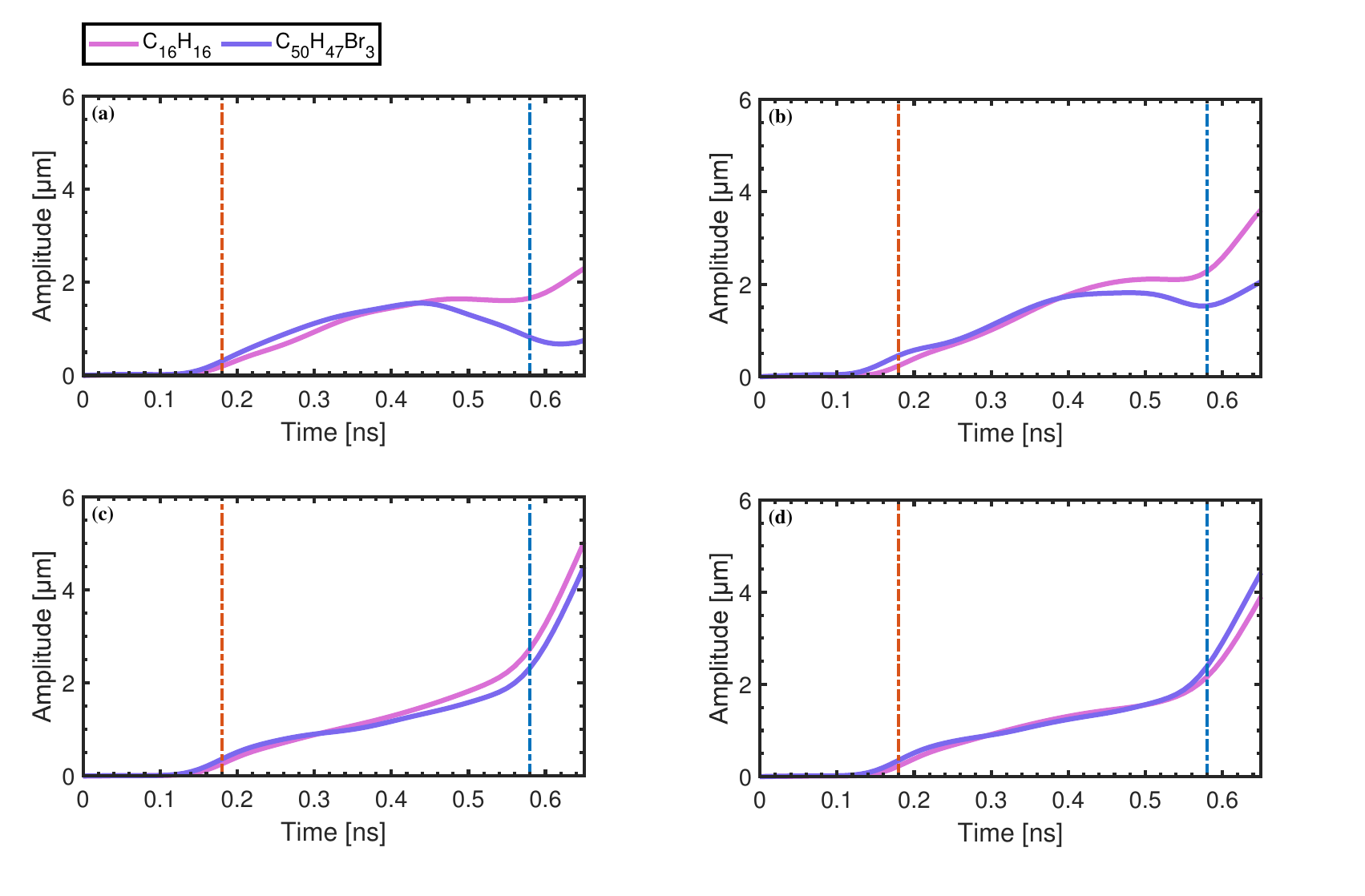}%
		\caption{\label{figure6} (Color online) Time evolution of the rippled inner-interface-initiated instability, with wavelengths of 10 $\mu\rm m$(a), 15 $\mu\rm m$ (b), 50 $\mu\rm m$(c) and 100 $\mu\rm m$(d), respectively. The red and blue vertical dash-dotted lines are the boundary between the feedout and ARM instability stage, and the boundary between the ARM and ART instability stage, respectively.}
	\end{figure*}
	
	Additionally, the double ablation front (DAF) is a distinct structure that appears only in high-Z doped targets, serving as a key feature for evaluating the validity of simulation results for such targets.
	However, as shown in Fig. \ref{figure6}(a), no distinct formation of the DAF structure is observed.
	This can be attributed to that the formation of this structure requires a certain amount of time, during which the Boltzmann number ($\rm Bo=\frac{5}{2}\it Pv/\sigma T^{\rm4}$) must fall below a specific threshold.\cite{DafTheory1}
	According to the theory of DAF,\cite{DafTheory1} radiation preheating increases over time, while the $T_e$ gradient continuously decreases. Eventually, a plateau with nearly uniform $T_e$ and $\rho$ forms. In this plateau, radiation and electrons reach local thermodynamic equilibrium (LTE), where they exhibit the same $T_r$ and $T_e$. 
	Meanwhile, two temperature fronts, the electron ablation front (EAF) and the radiation ablation front (RAF), are formed at both ends of this plateau, namely the DAF structure. 
	Based on the simulation results from Ref. \onlinecite{Xiong}, which use parameters similar to those in this article, the DAF structure becomes clearly visible at about 1.2 ns. 
	Since the simulation time scale in this study is short (0.65 ns), the DAF structure has not yet formed.
	
	\section{Evolution of the Rippled Inner-Interface-Initiated Instability at Different Wavelengths}\label{sec5}
	
	In this section, we change the wavelength of the perturbation to 10, 15, 50, and 100 $\mu\rm m$, and investigate its evolution at different wavelengths. 
	Other parameters such as the target settings and laser parameters, are the same as those used in Sec.\ref{sec3}. 
	The effect of high-Z dopant on the rippled inner-interface-initiated instability at different wavelengths is also investigated, including its effect on the feedout seeds and subsequent growth rate.
	
	Figures \ref{figure6}(a)-(d) present the simulation results of amplitude evolution in both high-Z doped and undoped targets at wavelengths of 10, 15, 50, and 100 $\mu\rm m$. The magenta and purple lines represent the results for undoped and high-Z doped targets, respectively. The two vertical dotted lines are the boundaries between the feedout and ARM instability stages, and between the ARM and ART instability stages. 
	From Fig. \ref{figure6}, it is found that in both high-Z doped and undoped targets, the whole perturbation evolution at different wavelengths still consists of three stages, namely the feedout, the ARM instability and the ART instability. 
	Additionally, for short-wavelength perturbations, oscillation occurs during the ARM instability stage, as Figs \ref{figure6}(a)-(b) shown. 
	While for long-wavelength perturbations, this oscillation disappears, as shown in Figs. \ref{figure6}(c) and \ref{figure6}(d). 
	During the ART instability, it is found that the growth rate of short-wavelength perturbations is lower than that of long-wavelength perturbations.
	
\begin{figure}\suppressfloats\centering
		\includegraphics[width=8cm]{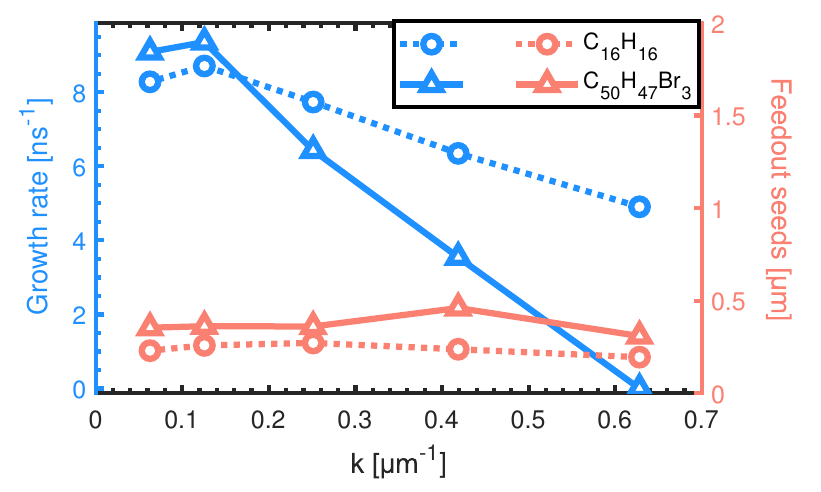}%
		\caption{\label{figure7} (Color online) The perturbation growth rate and feedout seeds at different perturbation wavelengths.}
	\end{figure}

	Fig. \ref{figure7} shows the growth rate and feedout seeds at different wavelengths. 
	The solid and dashed lines represent the results for the high-Z doped and undoped targets, respectively. 
	The growth rate, represented by the blue lines, is obtained by exponentially fitting the amplitude evolution curves during the ART instability stage in Figs. \ref{figure6}(a)-(d) and Fig. \ref{figure2.1}(e). 
	The feedout seeds are obtained by selecting the perturbation amplitude at the end of the feedout stage (approximately at 0.18 ns), represented by the orange lines in Fig. \ref{figure7}. 
	Specific values of feedout seeds and growth rates are also listed in Table \ref{table5}. 
	$\gamma^\star_1$ and $\gamma^\star_2$ represent the growth rate in the high-Z doped and undoped targets, respectively, while $x_{0,1}$ and $x_{0,2}$ denote the feedout seeds for the high-Z doped and undoped targets.
	
	As shown by the blue solid and blue dotted lines in Fig. \ref{figure7}, $\gamma^\star$ exhibits a characteristic of being large for long-wavelength perturbations and small for short-wavelength perturbations. 
	This can be attributed to the wavelength-dependent ablation stabilization effect.
	According to Eq. (\ref{eq-8.2}) and Eq. (\ref{eq-9.2}), the ablation stabilization is proportional to $k$. 
	Therefore, for short-wavelength perturbations (corresponding to large $k$ values), the ablation stabilization effect is significant, while for long-wavelength perturbations, it weakens. 
	
	In contrast, as shown by the orange dotted and solid lines in Fig. \ref{figure7}, the perturbation wavelength has little effect on the feedout seeds. 
	In our simulations, the long wavelength condition ($kd_{ps}<1$) is satisfied, allowing for analysis using the theory developed in Sec.\ref{sec2}, which demonstrates that the feedout seeds are independent of the perturbation wave number.
	For the high-Z doped targets, the feedout seeds is approximately 0.4 $\mu$m, while for the undoped targets, it ranges from 0.2 to 0.3 $\mu$m. These results are in good agreement with previous theoretical predictions in Table \ref{table1.3}.
	
	The effect of high-Z dopant on the rippled inner-interface-initiated instability at different wavelengths can be revealed by comparing the magenta and purple lines in Figs. \ref{figure6}. 
	During the feedout, the high-Z dopant slightly increases the feedout seeds, regardless of whether the perturbation wavelength is long or short. 
	Besides, during the ARM instability stage, it is found that the high-Z dopant decreases the ARM oscillation period and its amplitude, which is the same as previous findings\cite{XBBRM}. 
	Additionally, it is found that the high-Z dopant reduces the instability for perturbations with wavelengths of 10, 15, and 50 $\mu\rm m$ during the ART instability stage. 
	But for perturbations with a wavelength of 100 $\mu\rm m$, the high-Z dopant increases the ART instability.
	This suggests that for perturbations with short wavelengths, the high-Z dopant mitigates the rippled inner-interface-initiated ART instability, while for longer wavelengths, this instability is increased.
	
	\begin{table}
		\caption{\label{table5} Growth rate and feedout seeds at different perturbation wavelength in high-Z doped and undoped targets}
		\begin{ruledtabular}
			\begin{tabular}{ccccc}
				Wavelength & $\it\gamma^\star_{\rm1}$&$\it\gamma^\star_{\rm2}$&$\it x_{\rm{0,1}}$&$\it x_{\rm{0,2}}$\\
				($\mu \rm m$)&($\rm ns^{-1}$)&($\rm ns^{-1}$)&($\mu \rm m$)&($\mu \rm m$)\\
				\hline
				$10$&$0$\footnote{The perturbation of this wavelength is cut off.}&$4.90$&$0.31$&$0.19$\\
				$15$&$3.55$&$6.35$&$0.46$&$0.23$\\
				$25$&$6.43$&$7.73$&$0.36$&$0.27$\\
				$50$&$9.36$&$8.71$&$0.36$&$0.26$\\
				$100$&$8.28$&$9.09$&$0.35$&$0.23$\\
			\end{tabular}
		\end{ruledtabular}
	\end{table}

	Comparing the solid and dotted lines in Fig. \ref{figure7}, the effect of high-Z dopant at different wavelengths is more obvious. 
	For short-wavelengths perturbations($\lambda$ = 10, 15, 25 $\mu\rm m$), the growth rate in the high-Z doped targets is lower than that in the undoped targets, which is caused by the increased ablation velocity, as shown in Table \ref{table4}. 
	However, the difference in the ablation stabilization between the high-Z doped and undoped targets becomes negligible for long-wavelength perturbations ($\lambda$ = 100 $\mu\rm m$). 
	And the growth rate in the doped targets is almost identical to that in the undoped targets, with only a slight increase observed in the doped targets. 
	Additionally, the feedout seeds in the high-Z doped targets are always larger than that in the undoped targets, for both long wavelength and short wavelength perturbations.
	This is consistent with the theoretical predictions and analysis in previous sections, which is caused by the increased ablation pressure and decreased adiabatic exponent in the high-Z doped targets.
	
	Therefore, the influence mechanism of the high-Z dopant on the evolution of ARM and ART instabilities at different wavelengths can be revealed as follows. 
	The oscillation frequency of the ARM instability is proportional to the ablation velocity and the wave number. 
	For short wavelength perturbations in the high-Z doped targets, where both $v_a$ and $k$ are large, the ARM oscillation period is shorter than the ARM instability duration ($\delta t$), resulting the oscillation occurs within the ARM instability, as shown by purple lines in Figs. \ref{figure6}(a) and \ref{figure6}(b). 
	As perturbation wavelength increases, the ARM oscillation period decreases, then the oscillation disappear. 
	Due to lower $v_a$ in the undoped targets, the ARM oscillation occurs only for very short wavelength perturbations, where the wave number $k$ is sufficiently large, resulting in a short oscillation period.
	
	The perturbation during the ART instability stage depends on the feedout seeds and subsequent growth rate. 
	As orange solid and dotted lines shown in Fig. \ref{figure7}, for both long and short perturbation wavelengths, the feedout seeds are almost the same. 
	But there exists difference in the growth rate between the two targets for different perturbation wavelengths, as blue lines shown in Fig. \ref{figure7}.
	For short-wavelength perturbations, such as 10, 15, 25, and 50 $\mu\rm m$, although the feedout seeds in the high-Z doped targets are increased, the strong suppression of the high-Z dopant on the perturbation growth rate leads to a decrease in the final ART instability. This corresponds to cases in Figs. \ref{figure6}(a)-(c).
	For longer wavelength perturbations, such as 100 $\mu m$, the suppressing effect of the high-Z dopant weakens. 
	As a result, the increased feedout seeds in the high-Z doped targets play a major role in the subsequent perturbation evolution, causing the ART instability in high-Z doped targets to be larger than that in the undoped targets. This corresponds to case in Fig. \ref{figure6}(d).

	\section{Conclusion}\label{sec6}
	A theoretical model has been developed for the rippled inner-interface-initiated Rayleigh-Taylor instability. 
	The theory indicates that the evolution of this instability consists of three stages, namely the feedout, the ARM instability and the ART instability.
	Two-dimensional radiation hydrodynamic simulations provide validation for our theoretical model. The simulated feedout seeds show good agreement with the theoretical predictions.
	
	It is found that the high-Z dopant can increase the feedout seeds and decrease the subsequent growth.
	The increased feedout seeds in the doped targets can be attributed to their higher ionization, which results in a higher laser absorption rate, a higher critical surface density, and an increased isochoric molar heat capacity, both of which contributes to the increased feedout seeds.
	However, the growth rate of the ARM and ART instabilities is effectively suppressed, due to the X-ray preheat effect in the high-Z doped targets, resulting in the final perturbations is still decreased.
	
	Further research on this instability is conducted at different wavelengths perturbations.
	The feedout seeds are increased by the high-Z dopant, and are weakly dependent on the wavelength.
	But the ablation stabilization of subsequent growth is wavelength-dependent. 
	For short-wavelength perturbations, the high-Z dopant can still suppress the instability. 
	But for long-wavelength perturbations, the ablation stabilization weakens, leading to an increase in the final perturbations. 
	The results should be beneficial for the target designing to control the growth of ARM and ART instabilities in ICF.

	\begin{acknowledgments}\suppressfloats
		This work was supported by the National Natural Science Foundation of China (Grant Nos. 12175309, 12475252 and 12275356), the Strategic Priority Research Program of Chinese Academy of Science (Grant No. XDA25050200), the Defense Industrial Technology Development Program (Grant. JCKYS2023212807), Natural Science Foundation of Hunan Province, China (Grant No. 2025JJ20007), and the Postgraduate Scientific Research Innovation Project of Hunan Province, China (Grant No. CX20230005).
		
	\end{acknowledgments}

\end{document}